\documentstyle[aps,prb,preprint]{revtex}
\tightenlines

\begin{document}
\title{Long-Time Tails and Anomalous Slowing Down in the Relaxation of Spatially
Inhomogeneous Excitations in Quantum Spin Chains }
\author{G. O. Berim}
\address{R.O.S.E. Informathik GmbH,\\
Schlossstrasse 34, D-89518 Heidenheim, Germany,}
\author{S. Berim}
\address{Georg-Simon-Ohm-Fachhochschule Nuernberg,\\
Postfach 9012,Nuernberg, Germany,\\
and}
\author{G. G. Cabrera\thanks{%
e-mail: cabrera@ifi.unicamp.br}}
\address{Instituto de F\'{\i}sica ``Gleb Wataghin'',\\
Universidade Estadual de Campinas (UNICAMP), \\
C. P. 6165, Campinas 13083-970, SP, Brazil}
\date{Received May 28{\em th}, 2001; revised manuscript received September 16{\em %
th}, 2001}
\maketitle

\begin{abstract}
Exact analytic calculations in spin-1/2 $XY$ chains, show the presence of
long-time tails in the asymptotic dynamics of spatially inhomogeneous
excitations. The decay of inhomogeneities, for $t\rightarrow \infty $, is
given in the form of a power law 
\[
\left( t/\tau _{Q}\right) ^{-\nu _{Q}}\ ,
\]
where the relaxation time $\tau _{Q}$ and the exponent $\nu _{Q}$ depend on
the wave vector $Q$, characterizing the spatial modulation of the initial
excitation. We consider several variants of the $XY$ model (dimerized, with
staggered magnetic field, with bond alternation, and with isotropic and
uniform interactions), that are grouped into two families, whether the
energy spectrum has a gap or not. Once the initial condition is given, the
non-equilibrium problem for the magnetization is solved in closed form,
without any other assumption. The long-time behavior for $t\rightarrow
\infty $ can be obtained systematically in a form of an asymptotic series
through the stationary phase method. We found that gapped models show
critical behavior with respect to $Q$, in the sense that there exist
critical values $Q_{c}$, where the relaxation time $\tau _{Q}$ diverges and
the exponent $\nu _{Q}$ changes discontinuously. At those points, a slowing
down of the relaxation process is induced, similarly to phenomena occurring
near phase transitions. Long-lived excitations are identified as
incommensurate spin density waves that emerge in systems undergoing the
Peierls transition. In contrast, gapless models do not present the above
anomalies as a function of the wave vector $Q$.
\end{abstract}


\narrowtext

\section{Introduction}

The spin-$1/2$ $XY$ chain and its variants are among the most widely used
and quoted models in theoretical investigations on spin systems. This
popularity among theorists is due to two facts. On the one hand, this family
of models allows for exact theoretical description of many static as well as
dynamic properties(see Ref. \onlinecite{Lieb}-\onlinecite{Berim92}), and on
the other hand, in some important cases, it provides a good description of
real systems \cite{Larry,Harrison,TaylorD,basak}. As an example, we note
that the dimerized chain, with alternating antiferromagnetic bonds, has been
studied in relation with the spin-Peierls transition\cite{peierls}, and it
is thought to represent the spin degrees of freedom of organic compounds
that undergo the so called {\em Peierls distortion}. The present state of
the art in fabricating low dimensional systems, with the material science
technology developed after the synthesis of the superconducting cuprate
oxides, may now tailor compounds that reveal a wealth of new magnetic
phenomena, including random spin chains, spin ladders, and doped magnetic
chains among other systems with exotic behaviors. For theorists, these
systems are fascinating, with no parallel in classical or three-dimensional
physics. Being systems of low dimensionality with low values of the spin,
they are dominated by quantum effects\cite{dagotto}. Compounds such as $%
PrCl_{3}$, $PrEtSO_{4}$ ($Pr$ {\em Ethyl Sulfate}), and $Cs_{2}CoCl_{4}$ are
among the quasi-one-dimensional systems, whose low temperature properties
are thought to be described by the $XY$ model.

A great deal of theoretical information has been gathered concerning
equilibrium properties, with calculations of quantities such as the specific
heat, the magnetic susceptibility, and equal-time spin-spin correlation
functions\cite{Lieb,Niemeijer,Nishimori,Okamoto}.

As for the dynamic properties, the quantities most thoroughly studied are
the time-dependent spin-spin correlation functions (TDCF) 
\begin{equation}
\left\langle S_{j}^{\mu }(t)S_{l}^{\mu }\right\rangle =\left\langle \exp
\left( \frac{i}{\hbar }Ht\right) S_{j}^{\mu }\exp \left( -\frac{i}{\hbar }%
Ht\right) S_{l}^{\mu }\right\rangle \ ,  \label{cf}
\end{equation}
where $H$ is the Hamiltonian of the system, $\mu $ is an index for the spin
component ($\mu =x,y,z$), and $<...>$ is the equilibrium average\cite
{Niemeijer,McCoy71b,Sur,Stolze}. They are important for the description of
such dynamic phenomena as magnetic resonance, magnetic neutron scattering,
spin diffusion, and other relaxation properties\cite{Jongh,Steiner}. It
should be noted however, that their application is restricted to situations
near the equilibrium state, where the linear response theory is valid\cite
{Tomita,Kubo}. In particular, for the uniform $XY$ model, an old calculation
for the time dependent autocorrelation $<S_{0}^{z}(t)S_{0}^{z}(0\dot{)}>$
showed the absence of spin diffusion in the limit $t\rightarrow \infty $ 
\cite{Niemeijer}. This behavior was thought to be accidental and specific of
the $XY$ model for spin 1/2 (see, for example, Ref. \onlinecite{Steiner}).
However, recent research has shown that this surprising property is shared
by a whole family of integrable models, and is attributed to the existence
of a macroscopic number of conservation laws\cite{integral}. This remarkable
result has been probed in magnetic resonance experiments in some 1D systems%
\cite{TaylorD}. We will turn to this point later.

In spite of the wide range of their applicability, we note that TDCF do not
give a direct description of the time evolution of the non-equilibrium
process\cite{Jongh,Kubo}. This description is achieved through a different
method, that we briefly discuss below.

This second direction in investigating the dynamics relies on calculation of
averages of the type: 
\begin{equation}
\left\langle \ A\ \right\rangle _{t}=Tr\ [\rho (t)A]~,  \label{ave}
\end{equation}
where $\rho (t)$ is the density matrix which satisfies Liouville equation 
\begin{equation}
i\hbar \frac{\partial }{\partial t}\rho (t)=[H,\rho (t)]\ .
\label{liouville}
\end{equation}
Averages of such a kind give a direct account of the non-equilibrium
evolution of the physical observable $A$, independently of how far the
initial state is from the equilibrium state or from a stationary one.
Unfortunately, calculations of these quantities are much more involved than
the calculation of TDCF (\ref{cf}) for linear response, and few exact
results are known \cite
{Niemeijer,McCoy70,McCoy71a,Tjon,Vianen,Gibberd,Berim,Zaslavskii,Berim84,Berim91,Berim92}%
.

Whether the dynamical process is near or far from equilibrium, most works in
the literature deal with cases where the initial state is {\em spatially
uniform}. This premise is assumed in explicit or non-explicit way, and the
methods developed to solve the problem are heavily based on it. In contrast,
the study of spatially nonuniform excitations is practically new, in spite
of its interest, both theoretical and experimental. This problem is
important for a deeper understanding of dynamical processes in many-particle
spin systems with strong exchange interactions. Inhomogeneous initial states
can be prepared in real systems by external actions, for instance, strong
inhomogeneous magnetic fields or acoustic waves. On the other hand, from the
theoretical side, exact results on the dynamics (as in the case of the $XY$
model), can elucidate details of spin-spin relaxation processes in more
complicate systems.

In the present contribution, we will adopt the method based on formulae (\ref
{ave}) and (\ref{liouville}), and will analyze in detail the long time
evolution of the magnetization in three versions of the $XY$ model. The
initial excitation is always prepared in the form of a Spatially
Inhomogeneous Magnetization (SIM), and the calculation is done in exact
analytic form. The three variants of the $XY$ model that we consider are
enumerated below:

\begin{enumerate}
\item[I]  Isotropic dimerized $XY$ model with Hamiltonian: 
\begin{eqnarray}
H_{1}=-\mu
_{B}gh\sum_{j=1}^{N}~S_{j}^{z}-%
\sum_{m=1}^{M}~[J_{1}(S_{2m-1}^{x}S_{2m}^{x}+S_{2m-1}^{y}S_{2m}^{y})+ 
\nonumber \\
&&  \label{dimer} \\
\qquad \qquad \qquad +J_{2}(S_{2m}^{x}S_{2m+1}^{x}+S_{2m}^{y}S_{2m+1}^{y})];
\nonumber
\end{eqnarray}

\item[II]  Isotropic $XY$ model in a staggered magnetic field with
Hamiltonian:
\end{enumerate}

\begin{equation}
H_{2}=-\mu
_{B}g\sum_{m=1}^{M}(h_{1}S_{2m-1}^{z}+h_{2}S_{2m}^{z})-J%
\sum_{j=1}^{N}(S_{j}^{x}S_{j+1}^{x}+S_{j}^{y}S_{j+1}^{y});  \label{stagg}
\end{equation}

\begin{enumerate}
\item[III]  $XY$ model with bond alternation without magnetic field: 
\begin{equation}
H_{3}=-%
\sum_{m=1}^{M}(I_{1}S_{2m-1}^{x}S_{2m}^{x}+I_{2}S_{2m-1}^{y}S_{2m}^{y}+I_{2}S_{2m}^{x}S_{2m+1}^{x}+I_{1}S_{2m}^{y}S_{2m+1}^{y}).
\label{bond}
\end{equation}
\end{enumerate}

We will refer to them as models I, II, III, respectively. In expressions (%
\ref{dimer})-(\ref{bond}), $N=2M,$\ $\mu _{B}$\ is the Bohr magneton, $g$ is
the gyromagnetic ratio, $h$,\ $h_{1}$,\ $h_{2}$\ are magnetic fields, $J$,\ $%
J_{1}$,\ $J_{2}$, \ $I_{1}$,\ $I_{2}$\ are exchange integrals, $(J_{1}\ge
J_{2}>0$, and\ $J>0$), and cyclic boundary conditions are assumed. The above
restriction on $J_{1}$,\ $J_{2}$\ and $J$\ is not important in our problem
and is introduced for convenience.

To investigate SIM dynamics in the above models, we will use a previously
developed method given in Ref.\onlinecite{Berim84}. This method was applied
earlier by one of the authors\cite{Berim91,Berim92}, to the same study for
the isotropic $XY$ chain in a homogeneous magnetic field, with Hamiltonian: 
\begin{equation}
H_{4}=-\mu
_{B}gh\sum_{j=1}^{N}~S_{j}^{z}-J%
\sum_{j=1}^{N}(S_{j}^{x}S_{j+1}^{x}+S_{j}^{y}S_{j+1}^{y})\ .  \label{XY}
\end{equation}
We will refer to this model as model IV subsequently. We remark that highly
nontrivial results were obtained in Ref.\onlinecite{Berim91,Berim92}. In
particular, it was shown that in the $t\to \infty $ limit, some of the
spatially inhomogeneous excitations do not disappear and are still
time-dependent. The time evolution of SIM's (and also their spatial
distribution) is probed through the computation of the Fourier components $%
\left\langle S_{Q}^{z}(t)\right\rangle $ of the magnetization as a function
of the $Q$-wave number. The calculation yields a relaxation process in the
form of a power law 
\begin{equation}
\left( \frac{t}{\tau }\right) ^{-\nu }\ ,  \label{power}
\end{equation}
where the exponent $\nu $ depends on the initial state. In Ref.%
\onlinecite{Berim92} it was also studied the anisotropic $XY$ model in the
limit of strong anisotropy. It was found that the exponent $\nu $ in (\ref
{power}) changes discontinuously at some critical values $Q_{c}$ of the wave
vector $Q$. The values $Q_{c}$ depend solely on the parameters of the
Hamiltonian and are not connected with the preparation of the initial state,
nor the particular component of the magnetization that is relaxing.
Moreover, in the limit $Q\rightarrow Q_{c}$, the relaxation time $\tau $ of (%
\ref{power}) diverges to $\tau \rightarrow \infty $, with the corresponding
slowing down of the process. This phenomenon is at variance with some
conventional views for the time evolution of physical quantities in many
particle systems, according to which all quantities must be temporal
independent in the limit $t\to \infty $ . The concept of spin temperature,
widely used in non-equilibrium magnetic phenomena \cite{Goldman}, is based
on the assumption that the spin-spin relaxation is much faster than the
final relaxation to the lattice degrees of freedom.

Model I was also preliminary investigated by the some of the authors, and
similar conclusions were attained for the dimerized $XY$ model\cite
{BerimCabrera}, in the sense that a critical-like slowing down of the
relaxation process takes place at special points $\pm Q_{c}$ of the $Q$%
-space. In other words, for $\mid Q\mid \rightarrow Q_{c}$, the inverse time
scale\ $\tau ^{-1}$ for some of the oscillating components of SIM, goes to
zero following the power law $\tau ^{-1}\sim \left| \mid Q\mid -Q_{c}\right| 
$. Such behavior is not surprising for the case $Q_{c}=0$, for which the
corresponding value\ of the total magnetization $S^{z}$\ is a constant of
motion, but is unusual for $Q_{c}\neq 0,$ where the corresponding Fourier
component $S_{Q}^{z}$ is not conserved.

We note that $Q=0$ is the only critical point for the uniform isotropic $XY$
model (model IV and limit $J_{1}=J_{2}$ for model I and $h_{1}=h_{2}$ for
model II ). The $t\to \infty $ behavior is dominated by one oscillating
component, and this component has no critical properties. The present paper
is devoted to elucidate this remarkable difference within a more general
context, by the extensive study of a whole family of models. We suggest that
the dissimilar behaviors are due to the presence of a gap in the spectrum of
the energy excitations, the uniform isotropic model being gapless.

Our paper is organized as follows: In Section II, we outline the main steps
in the analytic calculation. All the models are diagonalized by means of a
modified Jordan-Wigner transformation, which maps the spin model into an
equivalent fermion Hamiltonian. Then, the average $<S_{Q}^{z}>_{t}$ is
calculated in exact closed form, and its asymptotic behavior for $%
t\rightarrow \infty $ is obtained after taking the thermodynamic limit $%
N\rightarrow \infty $. In Section III, we give a detailed analysis of the
results, including all models and the long time behavior of the
magnetization. Section IV closes the paper with the final discussions.

\section{Main steps of solution\label{two}}

The quantity of our interest is the time-dependent $z$-component of SIM 
\begin{equation}
\left\langle \ S_{Q}^{z}\ \right\rangle _{t}~\equiv ~Tr\ \left[ \rho
(t)S_{Q}^{z}\right] ~,  \label{SzQav}
\end{equation}
where $\rho (t)$ is the density matrix of the system, and 
\begin{equation}
S_{Q}^{z}=\sum_{j=1}^{N}~S_{j}^{z}\exp (iQj)~,  \label{ft}
\end{equation}
is the Fourier transform of the magnetization. In (\ref{SzQav}), $Q$ is the
wave vector characterizing the spatial inhomogeneity of the initial state ($%
Q=2\pi n/N,$ $n=-N/2+1,...,N/2$). To calculate $<S_{Q}^{z}>_{t}$\ one can
use the identity:

\[
\left\langle \ S_{Q}^{z}\ \right\rangle _{t}=\left\langle \ S_{Q}^{z}(t)\
\right\rangle _{0}~\equiv ~Tr\ \left[ \rho (0)S_{Q}^{z}(t)\right] \ , 
\]
where $\rho (0)$ is initial density matrix, and $S_{Q}^{z}(t)$ is the spin
operator in the Heisenberg representation 
\[
S_{Q}^{z}(t)=\exp \left( \frac{i}{\hbar }Ht\right) S_{Q}^{z}\exp \left( -%
\frac{i}{\hbar }Ht\right) \ . 
\]
To obtain exact results for $<S_{Q}^{z}>_{t}$, for each of the models
considered, we follow the steps:

\begin{enumerate}
\item[{\bf i)}]  Diagonalization of the corresponding Hamiltonian using the
fermion representation of the Jordan-Wigner transformation. At the end, we
get a free-fermion system;

\item[{\bf ii)}]  Transformation of $S_{Q}^{z}$\ in terms of fermionic
operators to get the time dependent $S_{Q}^{z}(t)$;

\item[{\bf iii)}]  Averaging of $S_{Q}^{z}(t)$ with the initial density
matrix $\rho (0)$ .
\end{enumerate}

The last step depends on the form of initial matrix $\rho (0)$. As it was
shown in Ref. \onlinecite{Berim84}, exact solutions of this problem may be
obtained in the case when $\rho (0)$ is a functional of only one component
of the spin operator: 
\begin{equation}
\rho (0)=F({S^{\gamma }}),\ \ (\gamma =x,y,z).  \label{inirho}
\end{equation}
Let us note that such initial state can actually be prepared in real systems
at low temperature, with a strong nonhomogeneous magnetic field directed
along the coordinate axis $\gamma $.

\subsection{Diagonalization}

Methods for diagonalization of Hamiltonians (\ref{dimer})-(\ref{bond}) are
well known (see for example Ref. \onlinecite{Taylor,Saika,Okamoto}). Minor
differences are specific for the version of the model to be solved. In our
case we have employed the following procedure:

\begin{enumerate}
\item[a)]  Use of the Jordan-Wigner transformation to change from spin to
Fermi operators $\left( b_{j}^{\dagger },\ b_{j}\right) $ :

\begin{equation}
S_{j}^{x}=L_{j}(b_{j}^{\dagger }+b_{j})/2,\quad
S_{j}^{y}=L_{j}(b_{j}^{\dagger }-b_{j})/2i,\quad S_{j}^{z}=b_{j}^{\dagger
}b_{j}-1/2\ ,  \label{jw}
\end{equation}
where 
\[
L_{j}=\prod_{l=1}^{j-1}(2b_{l}^{\dagger }b_{l}-1),\quad L_{1}\equiv 1~,\quad
(L_{j}^{2}=1)~;
\]

\item[b)]  Introduction of two types of Fermi operators for even and odd
sites: 
\begin{equation}
c_{m}^{\alpha }=b_{2m}^{\alpha },\quad d_{m}^{\alpha }=b_{2m-1}^{\alpha
},\quad m=1,2,...,M,\quad \alpha =\pm 1\ ,  \label{oddeven}
\end{equation}
and of their Fourier transforms $c_{k}^{\alpha }$~, $d_{k}^{\alpha }$~ : 
\begin{equation}
c_{k}^{\alpha }=\frac{1}{\sqrt{M}}\sum_{m=1}^{M}\exp (-i\alpha
km)c_{m}^{\alpha },\quad d_{k}^{\alpha }=\frac{1}{\sqrt{M}}%
\sum_{m=1}^{M}\exp [-i\alpha k(m-1/2)]d_{m}^{\alpha },  \label{blochop}
\end{equation}
$k=2\pi m_{1}/M$, with $m_{1}=-M/2+1,...,M/2$\ , where we have adopted the
compact notation 
\begin{equation}
\begin{array}{l}
c^{-}=c\ ,\qquad c^{+}=c^{\dagger }\ , \\ 
d^{-}=d\ ,\qquad d^{+}=d^{\dagger }\ ;
\end{array}
\label{compact}
\end{equation}

\item[c)]  After the above steps, Hamiltonians (\ref{dimer})-(\ref{bond})
are transformed into quadratic forms in terms of operators $\left(
c_{k}^{\alpha },\ d_{k}^{\alpha }~\right) $. Diagonalization of these
quadratic Hamiltonians is standard, but depends on the specific model under
consideration. For Hamiltonians (\ref{dimer}) and (\ref{stagg}) it can be
done with the help of the Bogolyubov transformation given below: 
\begin{equation}
c_{k}^{\alpha }=u_{k}\eta _{k}^{\alpha }+v_{k}^{\alpha }\beta _{k}^{\alpha
}\ ,\qquad d_{k}^{\alpha }=v_{k}^{-\alpha }\eta _{k}^{\alpha }-u_{k}\beta
_{k}^{\alpha }\ ,\qquad   \label{bogo}
\end{equation}
where$\left( \eta _{k}^{\alpha },\ \beta _{k}^{\alpha }\right) $\ are new
Fermi operators, $v_{k}^{-\alpha }=(v_{k}^{\alpha })^{*},$ and$\ u_{k}$\ is
a real function of $k$~. We are using the same convention (\ref{compact})
for $\alpha =\pm $.

For Hamiltonian (\ref{dimer}), the functions $\left( u_{k},~v_{k}\right) $
are given by: 
\begin{equation}
u_{k}=\frac{1}{\sqrt{2}}~,\quad v_{k}^{\alpha }=\frac{1}{\sqrt{2}}\exp
(i\alpha \theta _{k})  \label{coeff}
\end{equation}
where 
\[
\tan \theta _{k}=\frac{1-\delta }{1+\delta }\tan \left( k/2\right) ~,\quad
\delta =J_{2}/J_{1}\leq 1~.
\]

For Hamiltonian (\ref{stagg}), one gets: 
\begin{equation}
u_{k}=[(P_{k}-\omega _{d})/2P_{k}]^{1/2}~,\quad v_{k}=[(P_{k}+\omega
_{d})/2P_{k}]^{1/2}\mbox{sign}(\cos k/2)  \label{coeffstagg}
\end{equation}
where $v_{k}^{\alpha }$ is real (independent of $\alpha $) and 
\begin{equation}
\omega _{d}=\mu _{B}g(h_{1}-h_{2})/2\hbar ~,\quad P_{k}=(\omega
_{d}^{2}+\omega _{e}^{2}\cos ^{2}k/2)^{1/2}~,\quad \omega _{e}=J/\hbar ~.
\label{freq}
\end{equation}

\item[d)]  Following Ref. \onlinecite{Saika}, diagonalization of Hamiltonian
(\ref{bond}) is achieved through: 
\begin{equation}
c_{k}^{\alpha }=(-\eta _{k}^{\alpha }+\eta _{-k}^{\alpha }+\beta
_{k}^{\alpha }+\beta _{-k}^{\alpha })/2,\qquad d_{k}^{\alpha }=(\eta
_{k}^{\alpha }+\eta _{-k}^{\alpha }+\beta _{k}^{\alpha }-\beta _{-k}^{\alpha
})/2~.  \label{bogosp}
\end{equation}
\end{enumerate}

As a final result, all the three Hamiltonians can be represented in the
diagonal form 
\begin{equation}
\fbox{$
\begin{array}{c}
H_{i}=\hbar \sum_{k}\left[ \omega _{k}^{(i)}\eta _{k}^{\dagger }\eta
_{k}+\Theta _{k}^{(i)}\beta _{k}^{\dagger }\beta _{k}\right] \ ,\quad \\ 
(i=1,2,3)
\end{array}
$}  \label{diagonal}
\end{equation}
except for constant terms that are not important for the dynamics. The index 
$i=1,2,3$ refers to models I, II and III, respectively. The dispersion
relations of (\ref{diagonal}) are given by: 
\[
\begin{array}{c}
\omega _{k}^{(1)}=-\omega _{0}+\omega _{1}R_{k}~,\quad \omega
_{k}^{(2)}=-\omega _{s}+P_{k}~,\quad \omega _{k}^{(3)}=-(I_{2}/\hbar )\cos
\left( k/2\right) ~, \\ 
\\ 
\Theta _{k}^{(1)}=-\omega _{0}-\omega _{1}R_{k}~,\quad \Theta
_{k}^{(2)}=-\omega _{s}-P_{k}~,\quad \Theta _{k}^{(3)}=(I_{1}/\hbar )\cos
\left( k/2\right) ~,
\end{array}
\]
where 
\[
\begin{array}{l}
R_{k}=\sqrt{1+\delta ^{2}+2\delta \cos k}~,~~P_{k}=\sqrt{\omega
_{d}^{2}+\omega _{e}^{2}\cos ^{2}\left( k/2\right) }~, \\ 
\\ 
\omega _{0}=\mu _{B}gh/\hbar ~,\qquad \omega _{s}=\mu
_{B}g(h_{1}+h_{2})/2\hbar ~,\qquad \omega _{1}=J_{1}/2\hbar ~\ .
\end{array}
\]
Note that here and in the following, we neglect, as usually done\cite
{Lieb,McCoy70,Taylor,Okamoto}, the boundary term of the order $1/N$\ which
appeared in a chain with cyclic boundary conditions. The thermodynamic limit 
$N\rightarrow \infty $ will be taken at the end of the calculation.


\subsection{Calculation of $<S_{Q}^{z}>_{t}\label{calculation}$}

Using formulae (\ref{ft}) and (\ref{jw})-(\ref{blochop}), one can easily
obtain the following expression 
\begin{equation}
S_{Q}^{z}=-M\delta _{Q,0}+\sum_{k}(c_{p}^{+}c_{q}+d_{p}^{+}d_{q})\ ,
\label{sq}
\end{equation}
where $p=k-Q,\ q=k+Q$, with definitions common for all models. To obtain the
time dependence of $S_{Q}^{z}$, we first express the relation (\ref{sq}) in
terms of the canonical $\eta $'s and $\beta $'s through transformations (\ref
{bogo}) for models I and II, and (\ref{bogosp}) for model III. Then, we
insert the time evolution of such operators 
\[
\eta _{k}^{\alpha }(t)=\eta _{k}^{\alpha }\exp (i\alpha \omega
_{k}^{(i)}t),\quad \beta _{k}^{\alpha }(t)=\beta _{k}^{\alpha }\exp (i\alpha
\Theta _{k}^{(i)}t)~,\quad i=1,2 
\]
to get 
\begin{eqnarray}
S_{Q}^{z}(t) &=&-M\delta _{Q,0}+  \nonumber \\
+ &\sum_{k}&\left\{ (u_{p}u_{q}+v_{p}^{*}v_{q})\exp [i(\Omega
_{i}^{-}(k,Q)t]\eta _{p}^{+}\eta _{q}+(u_{p}u_{q}+v_{p}v_{q}^{*})\exp
[-i\Omega _{i}^{-}(k,Q)t]\beta _{p}^{+}\beta _{q}+\right.  \nonumber \\
&&\left. +(u_{p}v_{q}^{*}-u_{q}v_{p}^{*})\exp [i\Omega _{i}^{+}(k,Q)t]\eta
_{p}^{+}\beta _{q}+(u_{q}v_{p}-u_{p}v_{q})\exp [-i\Omega
_{i}^{+}(k,Q)t]\beta _{p}^{+}\eta _{q}\right\}  \label{Sqt}
\end{eqnarray}
for models I and II, where 
\[
\Omega _{i}^{+}(k,Q)=\omega _{p}^{(i)}-\Theta _{q}^{(i)}~,\qquad \Omega
_{i}^{-}(k,Q)=\omega _{p}^{(i)}-\omega _{q}^{(i)},\quad i=1,2 
\]
and functions $(u_{p},~v_{p})$ are defined by formulae (\ref{coeff}) and (%
\ref{coeffstagg}) respectively.

In the same way, for model III, we obtain: 
\begin{equation}
S_{Q}^{z}(t)=\sum_{k}\left\{ \beta _{-q}^{+}\eta _{p}^{+}\exp [i(\omega
_{p}^{(3)}+\Theta _{q}^{(3)})t]+\eta _{-p}\beta _{q}\exp [-i(\omega
_{p}^{(3)}+\Theta _{q}^{(3)})t]\right\} ~.  \label{SqtIII}
\end{equation}

To calculate the time dependent average $<S_{Q}^{z}>_{t}$, we use the
identity 
\[
\left\langle \ S_{Q}^{z}\ \right\rangle _{t}=\left\langle \ S_{Q}^{z}(t)\
\right\rangle _{0}~~, 
\]
which means that, equivalently, one can calculate the average of operator $%
S_{Q}^{z}(t)$ in the Heisenberg picture in relation to the initial density
matrix $\rho (0)$.

From equations (\ref{Sqt}) and (\ref{SqtIII}), one sees that this problem
reduces to calculation of averages of the type $<\eta _{p}^{\dagger }\eta
_{q}>_{0}$~, $<\beta _{p}^{\dagger }\beta _{q}>_{0}$~,..., including all the
combinations of canonical operators. The calculation is described in detail
in Appendix \ref{A}.

Below, we will summarize these results together with the ones obtained in
Ref. \onlinecite{Berim92} for model IV. We remark that exact solutions can
be obtained for special forms of the initial density matrix $\rho (0)$. We
assume here that $\rho (0)$ is a functional of only one component of the
spin operators,{\em \ i.e.}, $\rho (0)=F({S^{\gamma }}),\ \ (\gamma =x,y,z).$

\begin{enumerate}
\item  In the case $\gamma =x$ (the case $\gamma =y$ is the same by
symmetry) we have 
\begin{equation}
\left\langle \ S_{Q}^{z}\ \right\rangle _{t}=0  \label{null}
\end{equation}
for models I, III, IV, and 
\begin{eqnarray}
\left\langle \ S_{Q}^{z}\ \right\rangle _{t} &=&-\frac{i}{\pi }\omega
_{e}\omega _{d}\sin (Q/2)e^{iQ/2}\ \left\langle \ \varepsilon _{Q+\pi }^{x}\
\right\rangle _{0}  \nonumber \\
&&\int_{-\pi }^{\pi }\frac{\sin ^{2}\left( k/2\right) }{P_{p}P_{q}}\left\{
\cos [t\Omega _{2}^{+}(k,Q)]-\cos [t\Omega _{2}^{-}(k,Q)]\right\} dk,
\label{SQtx}
\end{eqnarray}
for model II, where 
\begin{equation}
\left\langle \ \varepsilon _{k}^{x}\ \right\rangle _{0}=\sum_{j=1}^{N}\
\left\langle S_{j}^{x}S_{j+1}^{x}\right\rangle _{0}\exp (ikj)~
\label{fourier}
\end{equation}
is the Fourier transform of the short-range order correlation of the $x$
component.

\item  In the case $\gamma =z$, we have different expressions for all the
cases.

\begin{enumerate}
\item  For model I, we get the formula: 
\begin{equation}
\left\langle \ S_{Q}^{z}\ \right\rangle _{t}=\frac{1}{4\pi }\int_{-\pi
}^{\pi }\left\{ F^{+}(k,Q)\cos [t\Omega _{1}^{+}(k,Q)]+F^{-}(k,Q)\cos
[t\Omega _{1}^{-}(k,Q)]\right\} dk,  \label{SQtzI}
\end{equation}
with 
\begin{equation}
\begin{array}{l}
F^{\pm }(k,Q)=A^{\pm }(k,Q)<S_{Q}^{z}>_{0}+B^{\pm }(k,Q)<S_{Q+\pi
}^{z}>_{0}\ , \\ 
A^{\pm }(k,Q)=1\mp 
{\displaystyle {(1+\delta ^{2})\cos Q+2\delta \cos k \over R_{p}R_{q}}}%
, \\ 
B^{\pm }(k,Q)=\pm 
{\displaystyle {i(1-\delta ^{2})\sin Q \over R_{p}R_{q}}}%
~.
\end{array}
\label{def}
\end{equation}

\item  For model II, we have the more explicit form: 
\begin{equation}
\begin{array}{l}
\left\langle \ S_{Q}^{z}\ \right\rangle _{t}=\frac{1}{4\pi }\left\langle \
S_{Q}^{z}\ \right\rangle _{0}\int_{-\pi }^{\pi }dk\  \\ 
\left\{ -\omega _{e}^{2}\cos (p/2)\cos (q/2)/P_{p}P_{q}\ \cos [t\Omega
_{2}^{+}(k,Q)]+\right.  \\ 
\left. +\left[ 1+\left( \omega _{d}^{2}+\omega _{e}^{2}\cos (p/2)\cos
(q/2)\right) /P_{p}P_{q}\right] \cos [t\Omega _{2}^{-}(k,Q)]~\right\} \ .
\end{array}
\label{SQtzII}
\end{equation}

\item  For model III, the compact expression: 
\begin{equation}
\left\langle \ S_{Q}^{z}\ \right\rangle _{t}=\frac{1}{2\pi }\left\langle \
S_{Q}^{z}\ \right\rangle _{0}\int_{-\pi }^{\pi }\cos [t\Omega (k,Q)]dk
\label{SQtzIII}
\end{equation}
where 
\begin{eqnarray}
\Omega (k,Q)=\upsilon _{Q}\cos (k/2+\alpha _{Q})\ ;  \nonumber \\
\tan \alpha _{Q}=\frac{I_{1}+I_{2}}{I_{1}-I_{2}}\tan (Q/2)~,\quad 
\label{defIII} \\
\hbar \upsilon _{Q}=\sqrt{I_{1}^{2}+I_{2}^{2}-2I_{1}I_{2}\cos Q}~.  \nonumber
\end{eqnarray}
Let us note that at the specific value $Q=\pi $, the time dependence of $%
\left\langle S_{Q}^{z}\right\rangle _{t}$ is given by the simple formula: 
\[
\left\langle \ S_{\pi }^{z}\ \right\rangle _{t}=\left\langle \ S_{\pi }^{z}\
\right\rangle _{0}\ J_{0}\left( \frac{I_{1}+I_{2}}{\hbar }t\right) ,
\]
where $J_{0}(x)$\ is the Bessel function of first kind.

\item  For model IV, it was found in Ref. \onlinecite{Berim92} 
\begin{equation}
\left\langle \ S_{Q}^{z}\ \right\rangle _{t}=\left\langle \ S_{Q}^{z}\
\right\rangle _{0}J_{0}(\Omega _{Q}t)~,  \label{SQtzIV}
\end{equation}
with $\Omega _{Q}=2(J/\hbar )\sin Q/2$~.
\end{enumerate}
\end{enumerate}



\section{Analysis of results\label{three}}

In this section, we firstly discuss general properties of the time evolution
of the magnetization, which can be inferred directly from equations (\ref{22}%
) and (\ref{null})-(\ref{SQtzIV}). In addition, we will study in detail the
asymptotic behavior at long times, in the limit $t\to \infty $. We will show
that the evolution of SIM's displays interesting unusual features, depending
on the wave vector $Q$ characterizing the inhomogeneous initial state.

\subsection{General properties of $<S_{Q}^{z}>_{t}\label{general}$}

Let us first consider model I in the case $\gamma =z$ for the initial
density matrix, since it presents a specific feature that is absent in the
other models studied here. From equations (\ref{22}) and (\ref{SQtzI})-(\ref
{def}), one sees that the value of the magnetization $<S_{Q}^{z}>_{t}$ at $%
t>0$, depends on the initial values of the two Fourier components $%
<S_{Q}^{z}>_{0}$ and $<S_{Q+\pi }^{z}>_{0}$. This means that, if only one
component of $S_{Q}^{z}$ exists at $t=0$, for example for wave vector $Q_{0}$%
, {\em i.e.} 
\[
<S_{Q}^{z}>_{0}=<S_{Q_{0}}^{z}>_{0}\delta _{Q,Q_{0}}~~, 
\]
then a twofold response develops for $t>0$, with components $%
<S_{Q_{0}}^{z}>_{t}$\ and\ $<S_{\pi +Q_{0}}^{z}>_{t}$, whose time evolution
is described by equation (\ref{22}), with $F^{\pm }(k,Q)=A^{\pm
}(k,Q_{0})<S_{Q_{0}}^{z}>_{0}\delta _{Q,Q_{0}}$ and $F^{\pm }(k,Q)=B^{\pm
}(k,Q_{0}+\pi )<S_{Q_{0}}^{z}>_{0}\delta _{Q,Q_{0}+\pi }$ , respectively. In
contrast, for models II-IV, if there is only one initial component $%
<S_{Q_{0}}^{z}>_{0}$, only one component $<S_{Q}^{z}>_{t}$ of the same wave
vector $Q=Q_{0}$ exists at subsequent times, $t>0$~.

When the initial condition is prepared with $\gamma =x$, the average $%
<S_{Q}^{z}>_{t}$ vanishes identically for models I, III and IV, for any $t>0$%
. This result may be understood intuitively, without any calculation, if one
notes that the part of the Hamiltonian that describes the coupling of spins
with the magnetic field, commutes with the other one describing the exchange
interactions (in the case of model III, this statement is trivial for we
only have the exchange part). Denoting the latter as $H_{ex}$, we get the
time evolution 
\begin{equation}
\left\langle \ S_{Q}^{z}\ \right\rangle _{t}=Tr\ \left[ \rho (0)\exp \left( 
\frac{i}{\hbar }H_{ex}\ t\right) S_{Q}^{z}\exp \left( -\frac{i}{\hbar }%
H_{ex}\ t\right) \right] ~,  \label{x1}
\end{equation}
which is governed by the exchange only. Inside the trace in (\ref{x1}), we
now perform a unitary transformation consisting in a rotation by $\pi $\
around the $x$\ axis. Under this operation, the spin operators transform as
follow: 
\[
S_{j}^{x}\to S_{j}^{x}~,~~S_{j}^{y}\to -S_{j}^{y}~,~~S_{j}^{z}\to
-S_{j}^{z}~. 
\]
The exchange Hamiltonian $H_{ex}$ is invariant under this transformation
because it is a sum of products of the form $S_{j}^{x}S_{j\pm 1}^{x}$ and $%
S_{j}^{y}S_{j\pm 1}^{y}$. The initial density matrix $\rho (0)$ is also
invariant since it only depends on the $S_{x}$ component of the spins ($%
\gamma =x$). So, the right hand side of equation (\ref{x1}) will change its
sign, yielding $<S_{Q}^{z}>_{t}=-<S_{Q}^{z}>_{t}$ i.e. $<S_{Q}^{z}>_{t}=0$\ .

In the case of model II, the exchange part does not commute with the total
Hamiltonian $H_{2}$. This leads, in general, to a nonvanishing $%
<S_{Q}^{z}>_{t}$, for $t>0$, which results proportional to $<\varepsilon
_{Q+\pi }^{x}>_{0}$ , where $<\varepsilon _{k}^{x}>_{0}\ $is the Fourier
transform of the spin-spin correlation function $<S_{j}^{x}S_{j+1}^{x}>_{0}\ 
$evaluated at the initial condition (see relation (\ref{fourier})). If the
initial state is prepared with a single $Q$-vector in the form $<\varepsilon
_{k}^{x}>_{0}=<\varepsilon _{Q_{0}}^{x}>_{0}\delta _{k,Q_{0}}$ , then $%
<S_{\pi +Q_{0}}^{z}>_{t}$ is the only component that exists for $t>0$\ .

For $Q=Q_{0}=0$ (homogeneous initial state), and for any $\gamma $, we get
the (evident) result: 
\[
<S_{Q=0}^{z}>_{t}=<S_{Q=0}^{z}>_{0} 
\]
for models I, II, IV. This is a direct consequence of the conservation of
the total $z$-component of the magnetization. In the case of model III, when 
$\gamma =z$, we get the closed result$:$%
\[
<S_{Q=0}^{z}>_{t}=<S_{Q=0}^{z}>_{0}J_{0}(\nu _{0}t) 
\]
where $\hbar \ \nu _{0}=\left| I_{1}-I_{2}\right| $\ .

Another important property valid for models I, II and IV, and for any $%
\gamma $, is that the evolution of SIM's is independent of the value of
external field $h$\ . This fact is not surprising because the Zeeman term of
the respective Hamiltonians commutes with both, the remaining part
describing the exchange interaction and the operator $S_{Q}^{z}$\ . This
way, the time evolution of SIM's is determined by the exchange term only.

Special important limits are $\delta =1\ (J_{1}=J_{2})$ in model I and $%
h_{1}=h_{2}$ in model II, that map onto model IV. In both of these cases,
the corresponding formulae for $<S_{Q}^{z}>_{t}$\ reduce to (\ref{SQtzIV})
which was obtained in Ref. \onlinecite{Berim92}.

Unfortunately, it is difficult to get more information from formulae (\ref
{22}), (\ref{SQtx}), (\ref{def}), and (\ref{SQtzII}), because they are
rather involved. So, in the following, we will consider the limiting case $%
t\rightarrow \infty $, that will be studied using the stationary phase
method. 

\subsection{Long time behavior of $<S_{Q}^{z}>_{t}$ for the initial
condition $\gamma =z$}

\subsubsection{Model I}

According to the stationary phase method \cite{statphase}, the long-time
evolution of $<S_{Q}^{z}>_{t}$ is dominated by the contribution of the
stationary points of the functions $\Omega _{1}^{\pm }(k,Q)$ . The number of
these points depends on the value of the wave vector $Q$. For convenience of
further discussion, let us introduce here the so called {\em critical}
values of $Q:$\newline
\begin{equation}
Q_{c_{1}}=\arccos \delta ,\quad Q_{c_{2}}=\pi -\arccos \delta ,\quad
Q_{c_{3}}=0.  \label{critical}
\end{equation}
The above values are determined in a standard manner through the stationary
phase method, being the locus where stationary points become degenerate\cite
{statphase}. Let us first examine the role of $\Omega _{1}^{+}(k,Q)$. It is
easy to show that it has five nondegenerate stationary points in the
interval $Q_{c_{1}}<\mid Q\mid <Q_{c_{2}}:$ 
\[
k_{1}=0,\qquad k_{2}=-k_{3}=\pi , 
\]
\[
k_{4}=-k_{5}=\varphi _{Q},\quad \varphi _{Q}=\arccos (-\delta ^{-1}\cos Q). 
\]

In turn, for $Q_{c_{2}}<\mid Q\mid \leq \pi $ and $Q_{c_{3}}<\mid Q\mid
<Q_{c_{1}}$, one gets three nondegenerate points at: 
\[
k_{1}=0,\quad k_{2}=-k_{3}=\pi . 
\]

Exactly at the critical values, we obtain:

\begin{enumerate}
\item[i)]  For $Q=\pm Q_{c_{1}}$, one nondegenerate point $k_{1}=0$ and two
degenerate ones $k_{2}=-k_{3}=\pi $ ($[\Omega _{1}^{+}(\pm \pi ,Q)]^{\prime
\prime }$ $=$ $[\Omega _{1}^{+}(\pm \pi ,Q)]^{\prime \prime \prime }=0,$\ $%
[\Omega _{1}^{+}(\pm \pi ,Q)]^{\left( IV\right) }\neq 0$ , where derivations
are taken with respect to $k$);

\item[ii)]  For $Q=\pm Q_{c_{2}}$ , two nondegenerate points $%
k_{2}=-k_{3}=\pi $, and one degenerate at $k=k_{1}=0\ ([\Omega
_{1}^{+}(0,Q)]^{\prime \prime }=[\Omega _{1}^{+}(0,Q)]^{\prime \prime \prime
}=0,[\Omega _{1}^{+}(0,Q)]^{IV}\neq 0)$;

\item[iii)]  For $Q=Q_{c_{3}}=0$ , there is no time evolution, since the
magnetization $<S_{Q=0}^{z}>_{t}$ is a constant of the motion, 
\[
\left\langle \ S_{Q=0}^{z}\ \right\rangle _{t}=\left\langle \ S_{Q=0}^{z}\
\right\rangle _{0}.
\]
\end{enumerate}

The function $\Omega _{1}^{-}(k,Q)$ vanishes at $Q=Q_{c_{3}}=0$\ and \ $%
Q=\pm \pi $. For any other wave number $Q$, it has two nondegenerate
stationary points: 
\[
k_{6}=-k_{7}=\psi _{Q},\quad \psi _{Q}=\arccos (-\delta \cos Q). 
\]
This finishes the analysis of the stationary points that dominate the
long-time behavior of (\ref{SQtzI}). The corresponding asymptotic
development of $<S_{Q}^{z}>_{t}$ for $t\rightarrow \infty $, can be
represented as a sum of several oscillating components with $Q$-dependent
frequencies and amplitudes: 
\begin{equation}
\left\langle \ S_{Q}^{z}\ \right\rangle _{t}\sim \sum_{i}S_{i}(Q,t),
\label{sum}
\end{equation}
\begin{equation}
S_{i}(Q,t)=\sum_{l=0}^{\infty }a_{i,l}(Q)\ \left( 
{\displaystyle {t \over \tau _{i}}}%
\right) ^{-\nu _{i}(2l+1)}\cos \left[ t\Omega _{i}(Q)+\alpha _{i,l}\right] \
,  \label{series}
\end{equation}
where $\tau _{i}\equiv \tau _{i}(Q)$ are functions of $Q$ and the exponent $%
\nu _{i}$ assumes the values $1/2$ or $1/4$ (the latter value will be
discussed in detail below). The number of components $S_{i}(Q,t)$ depends on 
$Q$. For $Q_{c_{1}}<\mid Q\mid <Q_{c_{2}}\ $(excluding $Q=\pm \pi /2$, which
is a degenerate point where $\Omega _{1}$ and $\Omega _{2}$ coincide), there
are four components. For $0\leq \mid Q\mid <Q_{c_{1}}$ and $Q_{c_{2}}<\mid
Q\mid <\pi $ (excluding accidental degeneracies), there are only three terms
in (\ref{sum}). The frequencies \ $\Omega _{i}(Q)$ and the inverse of the
characteristic times $\tau _{i}^{-1}$ of (\ref{series}) are given below: 
\begin{eqnarray}
\Omega _{1}(Q) &=&\omega _{1}\left( 1+\delta ^{2}+2\delta \cos Q\right)
^{1/2},  \nonumber \\
\tau _{1}^{-1} &=&\omega _{1}\delta \left( 1+\delta \cos Q\right) \ \frac{%
\left| \delta +\cos Q\right| }{R_{Q}^{3}},\quad Q\neq \pm Q_{c_{2}}\ , 
\nonumber \\
&&  \nonumber \\
\Omega _{2}(Q) &=&\omega _{1}\left( 1+\delta ^{2}-2\delta \cos Q\right)
^{1/2},  \nonumber \\
\tau _{2}^{-1} &=&\omega _{1}\delta \left( 1-\delta \cos Q\right) \ \frac{%
\mid \delta -\cos Q\mid }{R_{Q+\pi }^{3}},\quad Q\neq \pm Q_{c_{1}}\ , 
\nonumber \\
&&  \label{timesandfreq} \\
\Omega _{3}(Q) &=&\omega _{1}\left( \frac{R_{\psi _{Q}-Q}-R_{\psi _{Q}+Q}}{2}%
\right) \ ,  \nonumber \\
\tau _{3}^{-1} &=&\omega _{1}\frac{\delta }{2(1-\delta ^{2})}\left( R_{\psi
_{Q}-Q}+R_{\psi _{Q}+Q}\right) \mid \sin Q\mid \sin \psi _{Q},\ \ Q\neq
Q_{c_{3}}=0\ ,  \nonumber \\
&&  \nonumber \\
\Omega _{4}(Q) &=&\omega _{1}\left( \frac{R_{\varphi _{Q}-Q}+R_{\varphi
_{Q}+Q}}{2}\right) \ ,  \nonumber \\
\tau _{4}^{-1} &=&\omega _{1}\frac{\delta }{2(1-\delta ^{2})}\left(
R_{\varphi _{Q}-Q}-R_{\varphi _{Q}+Q}\right) \sin Q\sin \varphi _{Q},\ \
Q_{c_{1}}<\mid Q\mid <Q_{c_{2}}\ .  \nonumber
\end{eqnarray}
We display typical $Q$-dependences of $\Omega _{i}(Q)$ and $\tau _{i}^{-1}$%
in Figs.1 and 2, respectively, for particular values of $\delta $. Slowing
down of the relaxation process occurs at points where the $\tau _{i}^{-1}$'s
vanish.

At a generic $Q$-point, the first nonvanishing terms in (\ref{series}) have
the form: 
\begin{equation}
S_{i}(Q,t)\approx a_{i,0}(Q)\ \left( \frac{t}{\tau _{i}}\right) ^{-1/2}\cos
\left[ t\Omega _{i}(Q)+\alpha _{i,0}\right] \ ,\quad (i=1,...,4),  \label{36}
\end{equation}
with the amplitudes 
\[
a_{i,0}(Q)=a_{i,0}^{c}(Q)\ \left\langle \ S_{Q}^{z}\ \right\rangle
_{0}+ia_{i,0}^{s}(Q)\ \left\langle \ S_{Q+\pi }^{z}\ \right\rangle _{0}, 
\]
showing that the value at $Q$ depends on the two initial components $%
\left\langle S_{Q}^{z}\right\rangle _{0}$and $\left\langle S_{Q+\pi
}^{z}\right\rangle _{0}$. Formulae for the real and imaginary parts $%
a_{i,0}^{c}(Q)$ and $a_{i,0}^{s}(Q)$, as well as the phases $\alpha _{i,0}$,
are given in Appendix \ref{B}.

The dynamical process described by (\ref{sum}) and (\ref{series}), with the
explicit formulae given in (\ref{36}) and Appendix \ref{B}, is remarkable,
since it exhibits long time tails in the relaxation of SIM's to the
spatially homogeneous state. In contrast to the exponential relaxation,
which is characterized by a single parameter that yields the time scale or
the relaxation rate, the power law relaxation given by (\ref{series}) and (%
\ref{36}) is characterized by two parameters: $\tau _{i}^{-1}$, which
determines the inverse time scale of the process, and the exponent $\nu _{i}$%
, which determines the relaxation rate. In general, no conservation laws or
long-lived hydrodynamic modes seem to be associated with the above long time
tails.

In the neighborhood of critical points, the relaxation of some of the
components of (\ref{sum}) begins to stop, with the corresponding relaxation
time $\tau _{i}$ diverging, as shown in Fig. 2. Exactly in the limit, the
corresponding exponent $\nu _{i}$ jumps discontinuously from $1/2$ to $1/4$.
This slowing down of the relaxation process is very similar to the critical
slowing down found in phase transition phenomena.\cite{hohenberg}

Let us summarize below the singular behavior of the relaxation time at
special points (they are displayed in Fig. 2):

\begin{enumerate}
\item[{\em a)}]  $\tau _{1}^{-1}\rightarrow 0$ when $Q\rightarrow \pm
Q_{c_{2}}$, for any $\delta $ ;

\item[{\em b)}]  $\tau _{2}^{-1}\rightarrow 0$ when $Q\rightarrow \pm
Q_{c_{1}}$, for any $\delta $;

\item[{\em c)}]  $\tau _{4}^{-1}\rightarrow 0$ when $Q\rightarrow \pm
Q_{c_{1}},\ \pm Q_{c_{2}}$, for any $\delta $ ;

\item[{\em d)}]  $\tau _{3}^{-1}\rightarrow 0$ when $Q\rightarrow 0,\ \pm
\pi $, for any $\delta $.
\end{enumerate}

All the $\tau _{i}^{-1}$\ vanish according with the law: 
\[
\tau ^{-1}\sim \left| \left( \mid Q\mid -Q_{c}\right) \right| \ {\rm for}%
\mid Q\mid \rightarrow Q_{c}\ . 
\]
We now give the behavior of components of (\ref{sum}) at the critical points:

\begin{enumerate}
\item[I]  In the limit $Q\rightarrow \pm Q_{c_{1}}$, the components $%
S_{2}(Q,t)$ and $S_{4}(Q,t)$ merge into a single component with $\nu =1/4$: 
\begin{equation}
S_{2}(\pm Q_{c_{1}},t)=a_{2,0}(\pm Q_{c_{1}})\ \left( \frac{t}{\tau _{c}}%
\right) ^{-1/4}\cos \left[ t\Omega _{2}(Q_{c_{1}})+\pi /8\right] ,
\label{40}
\end{equation}
with 
\[
\begin{array}{l}
a_{2,0}(\pm Q_{c_{1}})=\frac{\sqrt[4]{24}}{8\pi }\Gamma (1/4)\left[
(1+\delta )\left\langle S_{\pm Q_{c_{1}}}^{z}\right\rangle _{0}+i\sqrt{%
1-\delta ^{2}}\left\langle S_{\pi \pm Q_{c_{1}}}^{z}\right\rangle
_{0}\right] \ , \\ 
\\ 
\tau _{c}^{-1}=3\omega _{1}\delta ^{2}/(1-\delta ^{2}),
\end{array}
\]
where $\Gamma (x)$ is the Gamma function. The other components follows the
law (\ref{36}) with $Q=\pm Q_{c_{1}}$;

\item[II]  Analogously, for $Q\rightarrow \pm Q_{c_{2}}$, the components $%
S_{1}(Q,t)$ and $S_{4}(Q,t)$ join together with $\nu =1/4$, and the time
dependence: 
\[
S_{1}(\pm Q_{c_{2}},t)=a_{1,0}(\pm Q_{c_{2}})\ \left( \frac{t}{\tau _{c}}%
\right) ^{-1/4}\cos \left[ t\Omega _{1}(Q_{c_{2}})+\pi /8\right] \ ,
\]
where 
\[
a_{1,0}(\pm Q_{c_{2}})=\frac{\sqrt[4]{24}}{8\pi }\Gamma (1/4)\left[
(1-\delta )\left\langle S_{\pm Q_{c_{2}}}^{z}\right\rangle _{0}+i\sqrt{%
1-\delta ^{2}}\left\langle S_{\pi \pm Q_{c_{2}}}^{z}\right\rangle
_{0}\right] \ .
\]
The other components follows the law (\ref{36}) with $Q=\pm Q_{c_{2}}$;

\item[III]  The component $S_{3}(Q,t)$ is critical at $Q=Q_{c_{3}}=0$, with
the relaxation time $\tau _{3}\sim 1/\left| Q\right| $ diverging in the
vecinity of $Q=0$. Note that the frequency $\Omega _{3}$ vanishes in the
limit $Q\rightarrow 0$. The criticality of $Q_{c_{3}}$ has a different
connotation, since it is related to the conservation of the total $z-$%
component of the spin 
\[
S_{Q=0}^{s}=\sum_{j=1}^{N}\ S_{j}^{z}\ ,
\]
which is valid for all the model with axial symmetry along the $z-$axis. The
divergence of $\tau _{3}$ near $Q\approx 0$ is connected to the stability of
hydrodynamic excitations of long wavelength;

\item[IV]  Note the exception of $Q\rightarrow \pm \pi $ for the component $%
S_{3}(Q,t)$, because the amplitude\ $a_{3,0}(Q)$\ goes to zero in this limit.
\end{enumerate}


\subsubsection{Model II}

When applying the same method to get the asymptotic behavior of (\ref{SQtzII}%
) for long times, one sees that the work is facilitated by the development
of the previous section. In fact, the frequencies $\Omega _{2}^{\pm }(k,Q)$\
and the functions $P_{k}$\ that appeared in (\ref{SQtzII}) of model II, can
be mapped onto the corresponding $\Omega _{1}^{\pm }(k,Q)$\ and $\omega
_{1}R_{k}$\ of model I, with the substitution: 
\begin{equation}
\delta \rightarrow \frac{\sqrt{\omega _{d}^{2}+\omega _{e}^{2}}-|\omega _{d}|%
}{\sqrt{\omega _{d}^{2}+\omega _{e}^{2}}+|\omega _{d}|}\ ,\qquad \omega
_{1}\rightarrow \frac{1}{2}{\sqrt{\omega _{d}^{2}+\omega _{e}^{2}}+}\frac{%
|\omega _{d}|}{2}.  \label{normalized}
\end{equation}
To avoid numerous definitions, we use the previous symbols $\delta $ and\ $%
\omega _{1}$ throughout this section, with the meaning given above in (\ref
{normalized}). With these variables, the stationary and critical points, the
inverse characteristic times $\tau _{i}^{-1}$\ and the frequencies $\Omega
_{i}(Q)$ coincide with those of model I.

The expression for $\left\langle S_{Q}^{z}\right\rangle _{t}$\ has the form (%
\ref{series}) with the following values of the first nonvanishing amplitudes 
$a_{i,k}(Q)$\ (at a generic $Q$-point): 
\begin{eqnarray*}
a_{1,1}(Q) &=&\left\langle \ S_{Q}^{z}\ \right\rangle _{0}\frac{\delta
(1-\delta )^{2}(1-\cos Q)}{R_{Q}^{4}\sqrt{8\pi }}\ , \\
a_{2,0}(Q) &=&\left\langle \ S_{Q}^{z}\ \right\rangle _{0}\ \sqrt{\frac{2}{%
\pi }}\ \frac{\delta (1-\cos Q)}{R_{Q+\pi }^{2}}\ , \\
a_{3,0}(Q) &=&\left\langle \ S_{Q}^{z}\ \right\rangle _{0}\ \sqrt{\frac{2}{%
\pi }}\ \frac{1+\delta \cos Q}{1+\delta }\ , \\
a_{4,0}(Q) &=&\left\langle \ S_{Q}^{z}\ \right\rangle _{0}\ \sqrt{\frac{2}{%
\pi }}\frac{\delta +\cos Q}{1+\delta }\ ,
\end{eqnarray*}
The corresponding phases $\alpha _{i,l}$\ are given by formulae (\ref{phases}%
). Note that the component with frequency $\Omega _{1}(Q)$\ decays to zero
as $t^{-3/2}$\ (exponent $\nu =1/2$ and $l=1$), i.e. more rapidly if
compared with the similar of model I . We now summarize the behavior at the
critical points:

\begin{enumerate}
\item[{\em i)}]  At the critical point $Q=\pm Q_{c_{1}}$, the components $%
S_{2}(Q,t)$\ and $S_{4}(Q,t)$\ are degenerate and decay to zero with
exponent $\nu =1/4$, following the law (\ref{40})\ with 
\[
a_{2,0}(\pm Q_{c1})=\frac{\sqrt[4]{24}}{2\pi }\Gamma (1/4)\frac{\delta }{%
1+\delta }~;
\]

\item[{\em ii)}]  At $Q=\pm Q_{c_{2}}$, the components $S_{1}(Q,t)$~ and $%
S_{4}(Q,t)$~ merge into a single component with the evolution ($\nu =1/4$
and $l=1$ in (\ref{40})) 
\begin{equation}
S_{1}(\pm Q_{c_{2}},t)=a_{1,1}(\pm Q_{c_{2}})\ \left( \frac{t}{\tau _{c}}%
\right) ^{-3/4}\cos \left[ t\Omega _{1}(Q_{c_{2}})+\pi /8\right] 
\label{40a}
\end{equation}
with 
\[
a_{1,1}(\pm Q_{c_{2}})=\frac{1}{2\pi }(3/2)^{3/4}\ \Gamma (3/4)\frac{\delta 
}{1+\delta }\ .
\]
\end{enumerate}

\subsubsection{Model III}

The long time behavior of (\ref{SQtzIII}) is much simpler than those
displayed by models I and II, for the phase function $\Omega (k,Q)$\ in (\ref
{defIII}) has only one stationary point at $k=k_{1}=-2\alpha _{Q}$\ . In the
asymptotic limit $t\to \infty $\ , it may be represented in the form 
\[
\left\langle \ S_{Q}^{z}\ \right\rangle _{t}=\left\langle \ S_{Q}^{z}\
\right\rangle _{0}\sqrt{\frac{2}{\pi }}(t/\tau _{Q})^{-1/2}\cos (t\ \upsilon
_{Q}-\pi /4) 
\]
where $\upsilon _{Q}$\ is given by formula (\ref{defIII}) and $\tau
_{Q}^{-1}=\upsilon _{Q}$~.



\subsection{Long time behavior of $\left\langle \ S_{Q}^{z}\ \right\rangle
_{t}$ for the initial condition $\gamma =x$}

As it was mentioned in Sec. \ref{calculation}, $\left\langle
S_{Q}^{z}\right\rangle _{t}$ , for the initial condition $\gamma =x$,
vanishes identically for models I, III, IV, and only has a nonzero value for
model II. For the latter, the asymptotic expression of $<S_{Q}^{z}>_{t}$\
for $t\to \infty $, has the form of (\ref{series}) with the quantities $%
\Omega _{i}$\ and $\tau _{i}$ defined through formulae (\ref{timesandfreq})
and (\ref{normalized}). The coefficients $a_{i,l}(Q)$\ for the first
nonvanishing amplitudes in (\ref{series}), at generic $Q$-points, are: 
\[
\begin{array}{l}
a_{1,1}(Q)=\left\langle \ \varepsilon _{Q+\pi }^{x}\ \right\rangle _{0}%
{\displaystyle {\sqrt{\delta }(1-\delta ) \over 2\sqrt{2\pi }}}%
(1-e^{iQ})/R_{Q}^{2}\ , \\ 
a_{2,0}(Q)=\left\langle \ \varepsilon _{Q+\pi }^{x}\ \right\rangle _{0}%
{\displaystyle {2\sqrt{\delta }(1-\delta ) \over \sqrt{2\pi }}}%
(1-e^{iQ})/R_{Q+\pi }^{2}\ , \\ 
a_{3,0}(Q)=-\left\langle \ \varepsilon _{Q+\pi }^{x}\ \right\rangle _{0}%
{\displaystyle {2\sqrt{\delta } \over \sqrt{2\pi }(1+\delta )}}%
(1-e^{iQ})(1+\delta \cos Q)\ , \\ 
a_{4,0}(Q)=\left\langle \ \varepsilon _{Q+\pi }^{x}\ \right\rangle _{0}%
{\displaystyle {2\sqrt{\delta } \over \sqrt{2\pi }(1+\delta )}}%
(1-e^{iQ})(1+\delta ^{-1}\cos Q)\ ,
\end{array}
\]
where $\left\langle \ \varepsilon _{k}^{x}\ \right\rangle _{0}\ $is the
correlation function defined in (\ref{fourier}). 
We remark that the first nonvanishing term for the component with frequency $%
\Omega _{1}$ has $\nu =1/2$ and $l=1$, yielding a $t^{-3/2}$ power law for
the time decay. The corresponding phases are given in (\ref{phases}) in
Appendix \ref{B}.

We make a summary of the critical behavior at singular points:

\begin{enumerate}
\item[{\bf i)}]  At the critical point $Q=\pm Q_{c_{1}}$, the components $%
S_{2}(Q,t)$\ and $S_{4}(Q,t)$\ merge into one component with time evolution
given by (\ref{40}), with $\nu =1/4$ and $l=0$, with the amplitude 
\[
a_{2,0}(\pm Q_{c_{1}})=\frac{\sqrt[4]{24}}{2\pi }\ \Gamma (1/4)\frac{\sqrt{%
\delta }}{1+\delta }\left( 1-\delta \mp i\sqrt{1-\delta ^{2}}\right) \ ;
\]

\item[{\bf ii)}]  In analogous form, at $Q=\pm Q_{c_{2}}$\ the components $%
S_{1}(Q,t)$\ and $S_{4}(Q,t)$\ become degenerate with the time evolution of (%
\ref{40a}), with $\nu =1/4$ and $l=1$, and the amplitude

\[
a_{1,1}(\pm Q_{c_{2}})=\frac{1}{2\pi }(3/2)^{3/4}\ \Gamma (3/4)\sqrt{\delta }%
\left( 1\mp i\sqrt{\frac{1-\delta }{1+\delta }}\right) \ .
\]
\end{enumerate}

\section{Final Discussion\label{four}}

We have studied the relaxation to the homogeneous state, of an initial
excitation which has been prepared with a nonhomogeneous magnetization
profile along the magnetic chain (SIM). The time evolution of this
excitation is probed through the calculation of the Fourier component $%
\left\langle S_{Q}^{z}(t)\right\rangle $ of the magnetization, which is
analyzed as a function of the wave vector $Q$. We use periodic boundary
conditions for the spin Hamiltonian and take the thermodynamic limit ($%
N\rightarrow \infty $) before studing asymptotic long times. As remarked in
Ref.\onlinecite{McCoy70}, the order of the limits is very important. All the
models treated here, show long time tails in the relaxation of $\left\langle
S_{Q}^{z}(t)\right\rangle $, which is apparent from the asymptotic study for 
$t\rightarrow \infty $, developed in the previous sections. This behavior
manifests itself in the form of a power law decay in the long time evolution
of SIM's. This slow relaxation is a remarkable result for its own sake and
is probably due to the absence of dissipation in the models. Since the
systems are isolated at zero temperature, the dynamics is exclusively driven
by quantum fluctuations.

Our calculation also shows striking differences, when one compares the
behaviors of models I and II on the one hand, with models III and IV, on the
other. The dissimilarity lies in the presence of critical values $Q_{c_{i}}$
for the wave vector $Q$ , where we get a slowing down of the relaxation
process. Near critical points, the time scale\ $\tau _{i}$ , giving the
relaxation of a specific oscillating component of $\left\langle
S_{Q}^{z}(t)\right\rangle $, diverges with the law $\tau ^{-1}\sim \left|
\mid Q\mid -Q_{c_{i}}\right| $, for $\mid Q\mid \rightarrow Q_{c_{i}}$,
reducing the damping of these excitations. Such behavior is not surprising
for the case $Q_{c_{3}}=0$, for which the corresponding value\ $S_{Q=0}^{z}$%
\ is an integral of the motion, but is unusual for the case $Q_{c_{i}}\neq 0,
$ when $S_{Q}^{z}$ is not conserved. In this sense, the criticality at $\pm
Q_{c_{1}}$, $\pm Q_{c_{2}}$ has different implications in the theory. The
position of the critical values $\pm Q_{c_{1}}$ and $\pm Q_{c_{2}}$ depends
only on parameters of the Hamiltonian, and in the limit $Q\rightarrow \pm
Q_{c_{1}},\pm Q_{c_{2}}$, the exponent $\nu $ jumps discontinuously, the
relaxation rate becoming slower than for excitations with $\left| Q\right|
\neq Q_{c_{1,2}}$. This means that the critical components will be the only
surviving ones at long enough times. 
These long-lived excitations may be pictured as spin density waves (charge
density waves for the fermion model) with incommensurate wave vector $%
Q_{c_{1,2}}$, which are developed in our gapped models as a consequence of
the Peierls instability\cite{peierls}. In contrast, the dynamics of models
III and IV does not have such anomalies. There is no gap in the spectrum,
and in the language of particles, the systems are always metallic and
display no critical points (except the point $Q=0$). In the asymptotic
regime $t\to \infty $ , only one oscillating component exists which displays
no critical behavior.

Peculiar properties of the energy spectrum seem to determine the critical
relaxation phenomena. In fact, models I and II display an energy gap between
the ground and excited states, whose size is monitored by the parameter $%
\delta $ (for model II through the transformation (\ref{normalized})). The
existence of critical points is directly related to the presence of the gap,
since the latter changes the curvature of the dispersion relation for the
energy. 
If one looks at formula (\ref{Sqt}) for the Fourier component $S_{Q}^{z}(t)$
of the magnetization, one realizes that processes which contribute with the $%
\pm \Omega _{i}^{+}$ frequency (the one that is critical) come from
transitions between both branches of the spectrum (interband transitions),
where a particle with momentum $q=k+Q$ is destroyed in a given band, and a
particle with momentum $p=k-Q$ is created in the other one, with a net
momentum transfer $\Delta k=\pm 2\left| Q\right| $. So the criticality at $%
Q_{c_{1,2}}$ is not directly related to the one particle spectrum, but to a
big density of states for those interband transitions that exchange the same
momentum $\Delta k=\pm 2\left| Q\right| $ coherently (constructive
interference). This{\em \ nesting} condition is attained at $Q=Q_{c_{1,2}}$,
with the onset of a spin density wave for the $S^{z}$ magnetization.

However, we have found that the above anomalous behavior in the relaxation
is not an exclusive property of the Peierls transition. In fact, other {\em %
1-dim} gapped spin models, as the $XY$ chain with anisotropic interactions,
present the same slowing down of the relaxation at special points (Ref.%
\onlinecite{Berim92} treats the Ising limit). A general connection between
this critical behavior and the spectral properties is still lacking and
currently under research\cite{tygel}.

Our results are all exact, with closed analytical forms, from where the
asymptotic regime for $t\rightarrow \infty $ is obtained. This has been done
using the fermion representation of the Jordan-Wigner approach for spin
chains. Other procedures to solve the same systems are available, the most
conspicuous being the Bethe Ansatz method, which applied to the $XY$ model
might hint at ``hidden'' conservation laws of the several versions of the
model treated here. However, it is difficult to conceive a similar analytic
calculation of the relaxation properties, using Bethe Ansatz techniques.

We remark that integrable models displays other unusual dynamical phenomena,
with anomalies in the transport properties\cite{integral}. For the
Heisenberg model, the absence of spin diffusion has been probed by
bosonization techniques\cite{narozhny} and numerical calculations\cite
{fabricius}. Ref. \onlinecite{fabricius} presents exact numerical
computations for the spin-1/2 $XXZ$ chain at $T=\infty $, which include, as
particular cases, the isotropic Heisenberg model and the $XY$ regime, which
are gapless, and the Heisenberg-Ising model, with a gap in the spectrum and
long-range order for the ground state. Numerical data is inconclusive due to
finite-size effects, but there are strong hints of a crossover from
non-diffusive to diffusive behavior when we go from the gapless region to
the gapped one. Concerning the low-temperature properties, a calculation by
Sachdev and Damle\cite{sachdev} in the gapped region, shows that the
diffusive behavior holds in the presence of long-range order. It is however
puzzling to admit that low energy properties of the spectrum may affect the $%
T=\infty $ behavior.

From a more fundamental point of view, we have studied the time evolution of
a closed quantum many body  system (at zero temperature), which is prepared
in an arbitrary nonhomogeneous initial state. The dynamics is solely given
by the Schr\"{o}dinger equation for the wave function or by the Liouville
equation for the density operator (which includes the more general situation
of a mixed ensemble for the initial state). Surprisingly, in the
thermodynamic limit and for long times ($N\rightarrow \infty $ first than
the limit $t\rightarrow \infty $), we get irreversibility in the form of a
power law relaxation for the magnetization, in spite that the dynamics is
unitary. No internal interactions are present, since all the Hamiltonians
are reduced to the free-particle form. The irreversibility can be entirely
ascribed to interference effects, that in our calculation are handled using
the stationary phase method at asymptotic long times. Due to constructive
interference, gapped models develop long-lived collective excitations with
the texture of spin density waves (charge density waves for the fermion
versions), that persist for longer times. We believe that those structures
are universal features of gapped one-dimensional models. 

Finally, our simple quantum systems are good candidates for {\em aging}
effects, a concept that has been coined to specify nonequilibrium dynamics
that depends on the initial conditions and relaxes very ``slowly''\cite
{aging}. But a complete characterization of quantum aging requires the
calculation of the two-times correlation functions\cite{jgc}.

\appendix

\section{Calculation of averages of canonical operators\label{A}}

We illustrate our method with the calculation of the averages $<\eta
_{p}^{\alpha }\eta _{q}^{\mu }>_{0}$, ($\alpha ,\mu =\pm $), for models I
and II. After taking the inverse of the Bogolyubov transformation 
\begin{equation}
\eta _{k}^{\alpha }=u_{k}c_{k}^{\alpha }+v_{k}^{\alpha }d_{k}^{\alpha
},\qquad \beta _{k}^{\alpha }=v_{k}^{-\alpha }c_{k}^{\alpha
}-u_{k}d_{k}^{\alpha }\ ,  \label{bogoinv}
\end{equation}
one obtains: 
\begin{eqnarray*}
<\eta _{p}^{\alpha }\eta _{q}^{\mu }>_{0} &=&u_{p}u_{q}<c_{p}^{\alpha
}c_{q}^{\mu }>_{0}+v_{p}^{\alpha }v_{q}^{\mu }<d_{p}^{\alpha }d_{q}^{\mu
}>_{0}+ \\
&+&u_{p}v_{q}^{\mu }<c_{p}^{\alpha }d_{q}^{\mu }>_{0}+v_{p}^{\alpha
}u_{q}<d_{p}^{\alpha }c_{q}^{\mu }>_{0}~.
\end{eqnarray*}
Then using definition (\ref{blochop}) and relations 
\[
c_{m}^{\alpha }=L_{2m}a_{2m}^{\alpha },\quad d_{m}^{\alpha
}=L_{2m-1}a_{2m-1}^{\alpha }, 
\]
where $a_{j}^{\alpha }\equiv S_{j}^{x}+i\alpha S_{j}^{y}~$ are the spin
ladder operators and $L_{j}=\prod_{l=1}^{j-1}(2S_{j}^{z})$ is the {\em string%
} operator of the Jordan-Wigner transformation\ (\ref{jw}), it is possible
to reduce the problem to calculation of the following spin averages: 
\begin{equation}
<a_{2j}^{\alpha }L_{2j}L_{2n}a_{2n}^{\mu }>_{0}~,\quad <a_{2j-1}^{\alpha
}L_{2j-1}L_{2n-1}a_{2n-1}^{\mu }>_{0}~,\quad <a_{2j}^{\alpha
}L_{2j}L_{2n-1}a_{2n-1}^{\mu }>_{0}~.  \label{spinave}
\end{equation}
All the averages above are taken with the initial density matrix $\rho (0)$.
If this is assumed to have the form (\ref{inirho}), then most averages that
involve products of different components of spin operators vanish. The only
nonzero averages are those with products of $\gamma -$components of spin
operators belonging to the same or neighboring lattice sites. Due to this
property, all averages in (\ref{spinave}) vanish except for $j=n,(n+1)$. The
results, for all the possible initial conditions $\rho (0)=F({S^{\gamma }})$%
, are given below:

\begin{enumerate}
\item[a)]  case $\gamma =x$ 
\begin{eqnarray*}
<a_{2j}^{\alpha }L_{2j}L_{2n}a_{2n}^{\mu }>_{0}=<a_{2j-1}^{\alpha
}L_{2j-1}L_{2n-1}a_{2n-1}^{\mu }>_{0}=\frac{1}{2}\delta _{j,n}\delta _{\mu
,-\alpha }~, \\
<a_{2j}^{\alpha }L_{2j}L_{2n-1}a_{2n-1}^{\mu }>_{0}=\mu \delta
_{n,j}<S_{2j}^{x}S_{2j-1}^{x}>_{0}-\alpha \delta
_{n,j+1}<S_{2j}^{x}S_{2j+1}^{x}>_{0}~;
\end{eqnarray*}

\item[b)]  case $\gamma =y$:
\end{enumerate}

\begin{eqnarray*}
<a_{2j}^{\alpha }L_{2j}L_{2n}a_{2n}^{\mu }>_{0} &=&<a_{2j-1}^{\alpha
}L_{2j-1}L_{2n-1}a_{2n-1}^{\mu }>_{0}=\frac{1}{2}\delta _{j,n}\delta _{\mu
,-\alpha }~, \\
<a_{2j}^{\alpha }L_{2j}L_{2n-1}a_{2n-1}^{\mu }>_{0} &=&-\alpha \delta
_{n,j}<S_{2j}^{y}S_{2j-1}^{y}>_{0}+\mu \delta
_{n,j+1}<S_{2j}^{y}S_{2j+1}^{y}>_{0}~;
\end{eqnarray*}

\begin{enumerate}
\item[c)]  case $\gamma =z$: 
\begin{eqnarray*}
<a_{2j}^{\alpha }L_{2j}L_{2n}a_{2n}^{\mu }>_{0}=\delta _{j,n}\delta _{\mu
,-\alpha }\left( \frac{1}{2}+\alpha <S_{2j}^{z}>_{0}\right) ~, \\
<a_{2j-1}^{\alpha }L_{2j-1}L_{2n-1}a_{2n-1}^{\mu }>_{0}=\delta _{j,n}\delta
_{\mu ,-\alpha }\left( \frac{1}{2}+\alpha <S_{2j-1}^{z}>_{0}\right) ~, \\
<a_{2j}^{\alpha }L_{2j}L_{2n-1}a_{2n-1}^{\mu }>_{0}=0~.
\end{eqnarray*}
\end{enumerate}

Next, one calculates all the averages contained in the expressions for $%
<S_{Q}^{z}(t)>_{0}$. We only give the results for $\gamma =x,z$ (the
remaining case $\gamma =y$ is identical with $\gamma =x$):

\begin{enumerate}
\item[A)]  Case $\gamma =x$~: 
\begin{equation}
\begin{array}{l}
<\eta _{p}^{+}\eta _{q}>_{0}=-(u_{p}v_{q}^{*}+u_{q}v_{p})<\varepsilon
^{x}(Q^{-})>_{0}-[u_{p}v_{q}^{*}e^{iq}+u_{q}v_{p}e^{-ip}]<\varepsilon
^{x}(Q^{+})>_{0}~, \\ 
<\beta _{p}^{+}\beta _{q}>_{0}=(u_{p}v_{q}+u_{q}v_{p}^{*})<\varepsilon
^{x}(Q^{-})>_{0}+[u_{p}v_{q}e^{-ip}+u_{q}v_{p}^{*}e^{iq}]<\varepsilon
^{x}(Q^{+})>_{0}, \\ 
<\eta _{p}^{+}\beta _{q}>_{0}=(u_{p}u_{q}-v_{p}v_{q})<\varepsilon
^{x}(Q^{-})>_{0}+[u_{p}u_{q}e^{iq}-v_{p}v_{q}e^{-ip}]<\varepsilon
^{x}(Q^{+})>_{0}~, \\ 
<\beta _{p}^{+}\eta _{q}>_{0}=(u_{p}u_{q}-v_{p}^{*}v_{q}^{*})<\varepsilon
^{x}(Q^{-})>_{0}+[u_{p}u_{q}e^{-ip}-v_{p}^{*}v_{q}^{*}e^{iq}]<\varepsilon
^{x}(Q^{+})>_{0}\ ,
\end{array}
\label{ave_x}
\end{equation}
where
\end{enumerate}

\[
\begin{array}{l}
<\varepsilon ^{x}(Q^{\pm })>_{0}=(<\varepsilon _{Q}^{x}>_{0}\pm <\varepsilon
_{Q+\pi }^{x}>_{0})/4M~,\quad \\ 
<\varepsilon _{k}^{x}>_{0}=\sum_{j=1}^{N}<S_{j}^{x}S_{j+1}^{x}>_{0}\exp
(ikj)~;
\end{array}
\]

\begin{enumerate}
\item[B)]  Case $\gamma =z$~:

\begin{equation}
\begin{array}{l}
<\eta _{p}^{+}\eta _{q}>_{0}=\frac{1}{2}\delta
_{p,q}+u_{p}u_{q}<S^{z}(Q^{+})>_{0}+v_{p}v_{q}^{*}<S^{z}(Q^{-})>_{0}\ , \\ 
<\beta _{p}^{+}\beta _{q}>_{0}=\frac{1}{2}\delta
_{p,q}+v_{p}^{*}v_{q}<S^{z}(Q^{+})>_{0}+u_{p}u_{q}<S^{z}(Q^{-})>_{0}~, \\ 
<\eta _{p}^{+}\beta
_{q}>_{0}=u_{p}v_{q}<S^{z}(Q^{+})>_{0}-u_{q}v_{p}<S^{z}(Q^{-})>_{0}~, \\ 
<\beta _{p}^{+}\eta
_{q}>_{0}=u_{p}v_{q}^{*}<S^{z}(Q^{+})>_{0}-u_{p}v_{q}^{*}<S^{z}(Q^{-})>_{0}~,
\end{array}
~  \label{21}
\end{equation}
where 
\[
<S^{z}(Q^{\pm })>_{0}=(<S_{Q}^{z}>_{0}\pm <S_{Q+\pi }^{z}>_{0})/2M~.
\]
\end{enumerate}

Using expressions (\ref{Sqt}) and (\ref{SqtIII}), together with (\ref{ave_x}%
) and (\ref{21}), and taking the continuum limit when $N\to \infty $, we get
the following expression for $<S_{Q}^{z}>_{t}$~: 
\begin{equation}
<S_{Q}^{z}>_{t}=\frac{1}{4\pi }\int_{-\pi }^{\pi }\{F^{+}(k,Q)\cos [t\Omega
_{i}^{+}(k,Q)]+F^{-}(k,Q)\cos [t\Omega _{i}^{-}(k,Q)]\}dk,  \label{22}
\end{equation}
where $i=1,\ 2$\ , for the two models treated here. Quantities in (\ref{22})
are 
\begin{eqnarray*}
F^{\pm }(k,Q)=\pm &&4u_{q}(u_{p}^{2}-\mid v_{p}\mid
^{2})[i(e^{-ip/2}+e^{iq/2})<\varepsilon _{Q}^{x}>_{0}\mbox{Im}\ v_{q} \\
&+&(e^{-ip/2}-e^{iq/2})<\varepsilon _{Q+\pi }^{x}>\mbox{Re}\ v_{q}]
\end{eqnarray*}
for $\gamma =x$~, and 
\begin{eqnarray*}
F^{\pm }(k,Q) &=&A^{\pm }(k,Q)<S_{Q}^{z}>_{0}+B^{\pm }(k,Q)<S_{Q+\pi
}^{z}>_{0}\ , \\
A^{\pm }(k,Q) &=&1\mp [1-2u_{p}^{2}u_{q}^{2}-2\mid v_{p}v_{q}\mid
^{2}-4u_{p}u_{q}\ \mbox{Re}\ (v_{p}^{*}v_{q})]~, \\
B^{\pm }(k,Q) &=&\pm 4iu_{p}u_{q}\ \mbox{Im}\ (v_{p}^{*}v_{q})
\end{eqnarray*}
for $\gamma =z$~.

Substituting the specific values (\ref{coeff}) and (\ref{coeffstagg}) for $%
(u_{p},v_{p})$ into formula (\ref{22}) one obtains the corresponding
expressions for models I and II given in Section II. In the same way,
analogous results may be obtained for model III.

\section{Amplitudes and phases\label{B}}

We give in this Appendix, formulae for the amplitudes and phases of the
oscillating components in the asymptotic development (\ref{series}), to
lowest order and for a generic $Q$-point :

\begin{equation}
S_{i}(Q,t)=a_{i,0}(Q)\ \left( \frac{t}{\tau _{i}}\right) ^{-1/2}\cos
[t\Omega _{i}(Q)+\alpha _{i,0}],\quad (i=1,...,4),
\end{equation}
with 
\begin{eqnarray}
a_{i,0}(Q) &=&a_{i,0}^{c}(Q)<S_{Q}^{z}>_{0}+ia_{i,0}^{s}(Q)<S_{Q+\pi
}^{z}>_{0},  \nonumber \\
&&  \nonumber \\
a_{1,0}^{c}(Q) &=&\frac{1}{\sqrt{8\pi }}(1-\delta )^{2}(1-\cos Q)/R_{Q}^{2},
\nonumber \\
a_{1,0}^{s}(Q) &=&\frac{1}{\sqrt{8\pi }}(1-\delta ^{2})\sin
Q/R_{Q}^{2},\quad Q\neq \pm Q_{c2};  \nonumber \\
&&  \nonumber \\
a_{2,0}^{c}(Q) &=&\frac{1}{\sqrt{8\pi }}(1+\delta )^{2}(1-\cos Q)/R_{Q+\pi
}^{2},  \label{amplitudes} \\
a_{2,0}^{s}(Q) &=&\frac{1}{\sqrt{8\pi }}(1-\delta ^{2})\sin Q/R_{Q+\pi
}^{2},\quad Q\neq \pm Q_{c1};  \nonumber \\
&&  \nonumber \\
a_{3,0}^{c}(Q) &=&\frac{1}{\sqrt{2\pi }}(1+\cos Q),\quad a_{3,0}^{s}(Q)=-%
\frac{1}{\sqrt{2\pi }}\sin Q,\quad Q\neq Q_{c3};  \nonumber \\
&&  \nonumber \\
a_{4,0}^{c}(Q) &=&\frac{1}{\sqrt{2\pi }}(1+\cos Q),\quad a_{4,0}^{s}(Q)=%
\frac{1}{\sqrt{2\pi }}\sin Q,\ \quad Q_{c1}<\mid Q\mid <Q_{c2}\ ;  \nonumber
\end{eqnarray}
and 
\begin{equation}
\alpha _{1,0}=\left\{ 
\begin{array}{cc}
-\pi /4, & \mid Q\mid <Q_{c2}, \\ 
\pi /4, & \mid Q\mid >Q_{c2},
\end{array}
\right. \qquad \alpha _{2,0}=\left\{ 
\begin{array}{cc}
-\pi /4, & \mid Q\mid >Q_{c1}, \\ 
\pi /4, & \mid Q\mid <Q_{c1},
\end{array}
\right.  \label{phases}
\end{equation}
\[
\alpha _{3,0}=-(\pi /4)\mbox{sign}\left( Q\right) ,\quad \alpha _{4,0}=\pi
/4\ . 
\]

\newpage

\begin{center}
{\bf FIGURE CAPTIONS}
\end{center}

{\bf Fig. 1 }Frequencies of the asymptotic components of the magnetization
in (\ref{series}), as functions of the wave number $Q$, for several values
of the parameter $\delta $. The different branches $\Omega _{i}$, for $%
i=1,2,3,4$, are indicated by the numbers. We also display the critical
points $\pm Q_{c_{1}}$ and $\pm Q_{c_{2}}$ by solid squares. Note that the $%
\Omega _{4}$ branch is tangent to the $\Omega _{1}$ and $\Omega _{2}$
branches at $Q_{c_{2}}$ and $Q_{c_{1}}$ respectively, but it is only defined
between critical points, $Q_{c_{1}}<\left| Q\right| <Q_{c_{2}}$. The
complete $\Omega _{4}$ curve is shown as a guide for the eye.

\medskip

{\bf Fig. 2 }Inverse of the relaxation times $\left( \omega _{1}\tau
_{i}\right) ^{-1}$ of the power-law decay of (\ref{series}), as functions of
the wave number $Q$ for the same values of the parameter $\delta $ displayed
in the previous figure. The numbers label the different branches, in close
correspondence with Fig. 1. Solid squares are used to display $\left( \omega
_{1}\tau _{4}\right) ^{-1}$, that vanishes at the critical points. The
arrows in the upper figure indicate values out of the scale.


\begin{references}
\bibitem{Lieb}  E. Lieb, T. Schultz and D. Mattis, Ann. Phys. (N.Y.) {\bf 16}%
, 406 (1961).

\bibitem{Katsura}  S. Katsura, Phys. Rev. {\bf 127}, 1508 (1962).

\bibitem{Niemeijer}  Th. Niemeijer, Physica {\bf 36}, 377 (1967); Physica 
{\bf 39}, 313 (1968).

\bibitem{McCoy70}  E. Barouch, B. M. McCoy and M. Dresden, Phys. Rev. A {\bf %
2}, 1075 (1970).

\bibitem{McCoy71}  E. Barouch and B. M. McCoy, Phys. Rev. A {\bf 3}, 2137
(1971).

\bibitem{McCoy71a}  E. Barouch and B. M. McCoy, Phys. Rev. A{\bf 3}, 786
(1971).

\bibitem{McCoy71b}  B. M. McCoy, E. Barouch and D. B. Abraham, Phys. Rev. A%
{\bf 4}, 2331 (1971).

\bibitem{Nishimori}  H. Nishimori, Phys. Lett. {\bf 100A}, 239 (1984).

\bibitem{Taylor}  J. H. Tailor and G. M{\"{u}}ller, Physica A {\bf 130}, 1
(1985).

\bibitem{Shrock}  G. M{\"{u}}ller and R. E. Shrock, Phys. Rev. B {\bf 31},
637 (1985).

\bibitem{Saika}  Y. Saika, J. Phys. Soc. Jpn. {\bf 63}, 3983 (1994).

\bibitem{Satija}  I. I. Satija and J. C. Chaves, Phys. Rev. B {\bf 49},
13239 (1994).

\bibitem{Okamoto}  K. Okamoto, J. Phys. Soc. Jpn. {\bf 57}, 2947 (1988); 
{\em ibid} {\bf 58}, 2004 (1989).

\bibitem{Sur}  A. Sur, D. Jasnow and I. J. Lowe, Phys. Rev. B {\bf 12}, 3845
(1975).

\bibitem{Brandt}  U. Brandt and K. Jacoby, Z. Physik B {\bf 25}, 181 (1976); 
{\em ibid} B {\bf 26}, 245 (1977).

\bibitem{Capel}  H. W. Capel and J. H. H. Perk, Physica A {\bf 87}, 211
(1977).

\bibitem{Perk}  J. H. H. Perk and H. W. Capel, Physica A {\bf 89}, 265
(1977); {\em ibid} {\bf 92}, 163 (1978); {\em ibid} {\bf 100}, 1 (1980).

\bibitem{Puga}  M. W. Puga and H. Beck, J. Phys. C:{\em Solid State} {\bf 15}%
, 2441 (1982).

\bibitem{Muller}  G. M\"{u}ller and R. E. Shrock, Phys. Rev. Lett. {\bf 51},
219-22 (1983).

\bibitem{Perk84}  J. H. H. Perk, H. W. Capel, G. R. W. Quispel and F. W.
Nijhoff, Physica A {\bf 123}, 1 (1984).

\bibitem{Apel}  W. Apel, Z. Phys B {\bf 63}, 185 (1986).

\bibitem{Its}  A. R. Its, A. G. Izergin, V. E. Korepin, and N. A. Slavnov,
Phys. Rev. Lett. {\bf 70}, 1704 (1993).

\bibitem{Stolze}  J. Stolze, A. N\"{o}ppert and G. M\"{u}ller, Phys.Rev. B 
{\bf 52}, 4319 (1995); M. B\"{o}hm, H. Leschke, Physica A {\bf 199}, 116
(1993); M. B\"{o}hm, H. Leschke, M. Henneke, V. S. Viswanath, J. Stolze, G.
M\"{u}ller, Phys. Rev. B {\bf 49}, 417 (1994); K. Fabricius, U. L\"{o}w, and
J. Stolze, Phys. Rev. B {\bf 55}, 5833 (1997).

\bibitem{Jongh}  L. J. de Jongh and A. R. Miedema, Adv. Phys. {\bf 23}, 1
(1974).

\bibitem{Steiner}  M. Steiner, J. Villain and C. G. Windsor, Adv. Phys. {\bf %
25}, 87 (1976).

\bibitem{Tomita}  R. Kubo and K. Tomita, J. Phys. Soc. Jpn. {\bf 9}, 888
(1954).

\bibitem{Kubo}  R. Kubo, J. Phys. Soc. Jpn. {\bf 12}, 570 (1957).

\bibitem{Tjon}  J. A. Tjon, Phys. Rev. B {\bf 2}, 2411 (1970).

\bibitem{Vianen}  H. A. W. van Vianen and J. A. Tjon, Physica {\bf 48}, 497
(1970).

\bibitem{Gibberd}  B. W. Gibberd and G. C. Stew, Phys. Lett. {\bf 36A}, 353
(1971).

\bibitem{Berim}  G. O. Berim and A. R. Kessel, Physica A {\bf 116}, 526
(1982); {\em ibid} {\bf 117}, 603 (1983); {\bf 119}, 153 (1983).

\bibitem{Zaslavskii}  O. B. Zaslavskii and V. M. Tsukernik, Sov. J. Low
Temp. Phys. {\bf 6}, 647 (1980); {\em ibid} {\bf 9}, 33 (1983).

\bibitem{Berim84}  G. O. Berim and A. R. Kessel, Teor. Mat. Fiz. {\bf 58},
388 (1984), in Russian.

\bibitem{Berim91}  G. O. Berim, Sov. J. Low Temp. Phys. {\bf 17}, 329 (1991).

\bibitem{Berim92}  G. O. Berim, Sov. Phys. JETP {\bf 75}, 86 (1992).

\bibitem{Larry}  L. W. Culvahouse, D. P. Schrinke, G. Larry and K.
Pfortmiller, Phys. Rev. {\bf 117}, 454 (1969).

\bibitem{Harrison}  J. P. Harrison, J. P. Heccler and D. P. Taylor, Phys.
Rev. B {\bf 14}, 2979 (1976).

\bibitem{TaylorD}  D. R. Taylor Phys. Rev. Lett. {\bf 42}, 1302 (1979).

\bibitem{basak}  R. Basak and I. Chatterjee, Phys. Rev. B {\bf 40}, 4627
(1989).

\bibitem{peierls}  S. Kagoshima, H. Nagasawa, and T. Sambongi, {\em %
One-Dimensional Conductors} (Springer-Verlag, Berlin, 1988).

\bibitem{dagotto}  E. Dagotto and T. M. Rice, Science {\bf 271}, 618 (1996).

\bibitem{integral}  B. M. McCoy, in {\em Statistical Mechanics and Field
Theory}, edited by V. V. Bazhanov and C. J. Burden, (World Scientific,
Singapore, 1995), p. 26; H. Castella, X. Zotos and P. Prelov\v {s}ek, Phys.
Rev. Lett. {\bf 74}, 972 (1995).

\bibitem{Goldman}  M.Goldman {\em Spin Temperature and Nuclear Magnetic
Resonance in Solids}, (Clarendon, Oxford, 1970).

\bibitem{BerimCabrera}  G. Berim and G. G. Cabrera, Physica A {\bf 238}, 211
(1997).

\bibitem{statphase}  M. V. Fedoryuk, {\em Asymptotic Behavior: Integrals and
Series} (Nauka, Moscow, 1984), in Russian; A. Erd\'{e}lyi, {\em Asymptotic
Expansions} (Dover, New York, 1956); J. J. Stamnes, {\em Waves in Focal
Regions} (Adam Hilger, Bristol and Boston, 1986).

\bibitem{hohenberg}  P. C. Hohenberg and B. I. Halperin, Rev. Mod. Phys. 
{\bf 49}, 435 (1977)

\bibitem{tygel}  D. Tygel and G. G. Cabrera, not published.

\bibitem{narozhny}  B. N. Narozhny, Phys. Rev. B {\bf 54}, 3311 (1996).

\bibitem{fabricius}  K. Fabricius and B. M. McCoy, Phys. Rev. B {\bf 57},
8340 (1998).

\bibitem{sachdev}  S. Sachdev and K. Damle, Phys. Rev. Lett. {\bf 78}, 943
(1997); K. Damle and S. Sachdev, Phys. Rev. B {\bf 57}, 8307 (1998).

\bibitem{aging}  L. F. Cugliandolo and G. Lozano, Phys. Rev. Lett. {\bf 80},
4979 (1998); G. M. Sch\"{u}tz and S. Trimper, Europhys. Lett. {\bf 47}, 164
(1999); F. Igl\'{o}i and H. Rieger, Phys. Rev. Lett. {\bf 85}, 3233 (2000);
N. Pottier and A. Mauger, Physica A {\bf 282}, 77 (2000).

\bibitem{jgc}  J. G. Carvalho, D. Tygel and G. G. Cabrera, work under
progress.
\end{references}
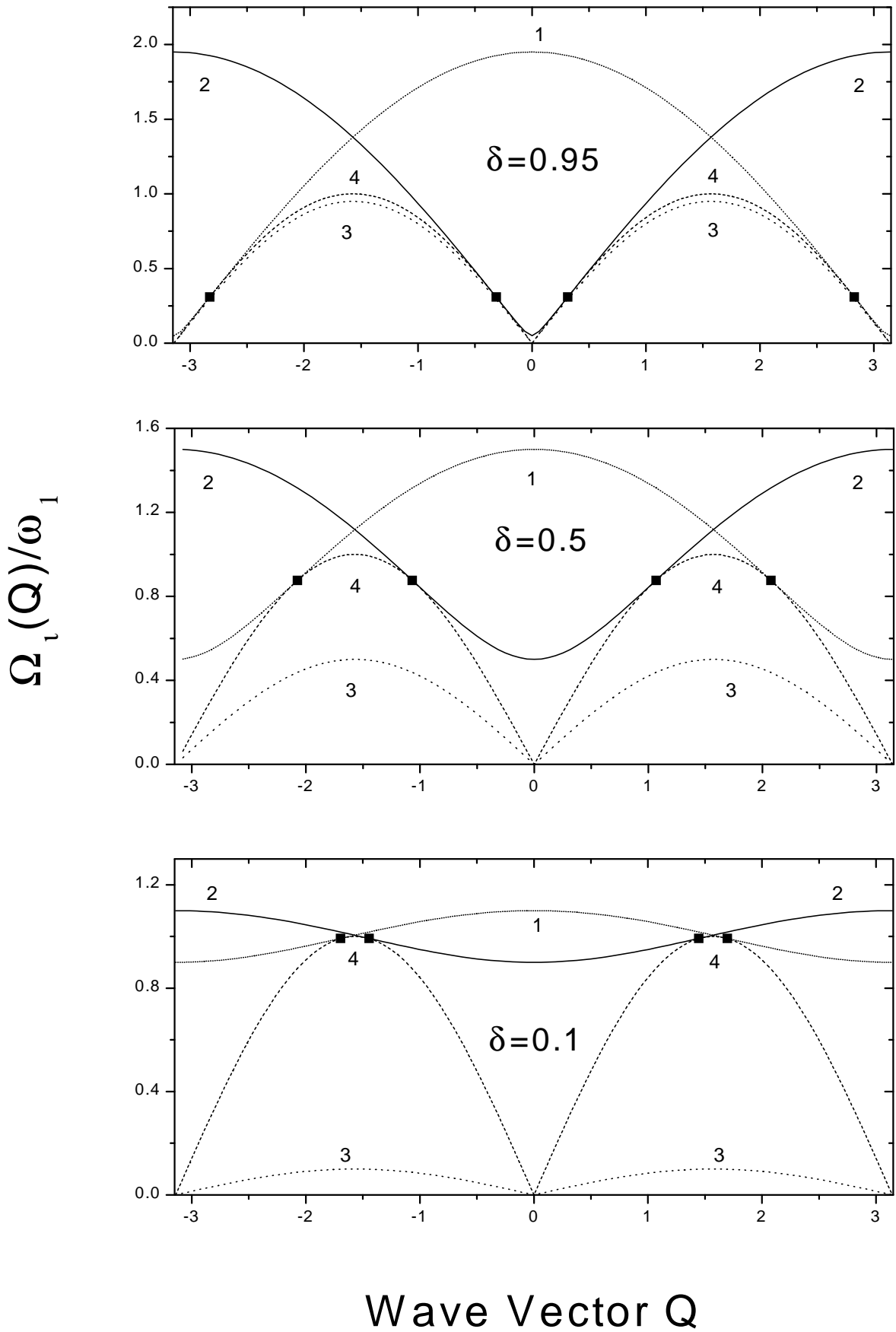
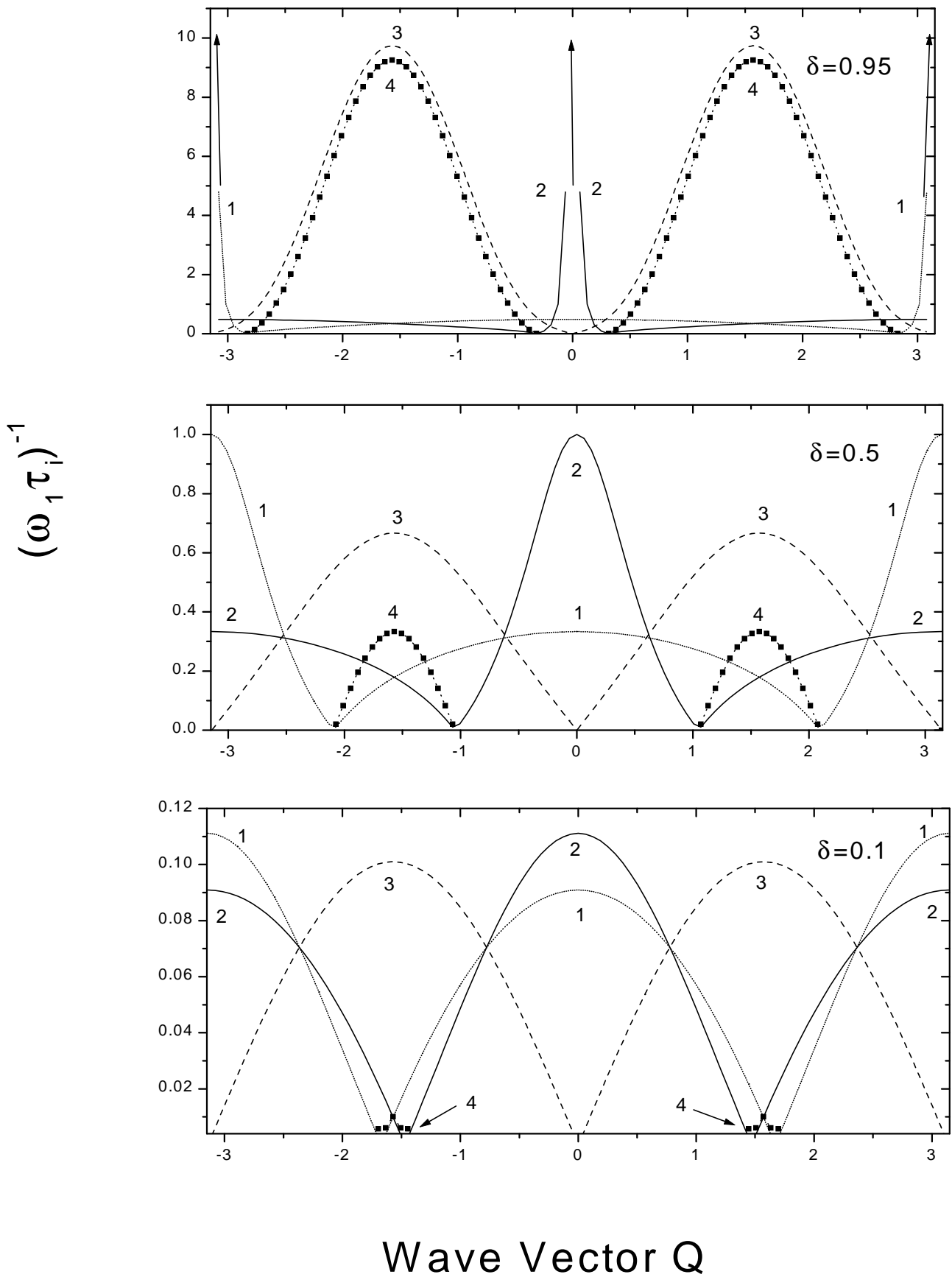
\end{document}